\documentstyle[preprint,aps]{revtex}
\begin{document}
\preprint {TR-PR-93-77}
\draft
\date{\today}
\title{The nuclear component of the EMC effect beyond the Plane
Wave Impulse Approximation}
\author{S. A. Gurvitz and A.S. Rinat}
\address{Department of Particle Physics, Weizmann Institute of
         Science, Rehovot 76100, Israel\\
and \\TRIUMF, Vancouver, B.C., Canada V6T-2A3}
\maketitle
\begin{abstract}
We study  the response  of a  nucleus composed  of nucleons  and confined
quarks.   Assuming dynamics implied by a  non-relativistic cluster
model, we prove that the total response is a convolution of the responses
of,  respectively,  a  nucleus  with interacting  point-nucleons  and  an
isolated  nucleon  with confined  quarks.   Defining  as an  intermediary
structure functions in terms of  the non-relativistic limit of light-cone
variables,   we  subsequently   conjecture   a   generalization  to   the
relativistic regime. That result contains a nuclear part with
full inter-nucleon dynamics and allows the study of  approximations, as
are  the Plane  Wave  Impulse  Approximation and  a  modification of  it.
The framework also permits a clean treatment of the response of
a moving composite nucleon in the nucleus which  may be off-shell.

\end{abstract}
\section{Introduction}

Consider  the cross  section for  deep inelastic  scattering of  a weakly
interacting  probe by  a nuclear  target.  Factoring  out the  elementary
cross section, one is left with  the nuclear structure function or linear
response, which has  been measured for various targets  $A$.  The outcome
of  the  result per  nucleon  differs  from  structure function  of  a
nucleon \cite{aub}.   In spite of  protracted efforts to  describe those
systematic  deviations--the  so-called EMC  effect--there  is  as yet  no
consensus on a satisfactory explanation \cite{btom}.

It is commonly accepted that at least the following two causes contribute
to the  EMC effect. 1) A  redistribution of quarks ($'\alpha'$)  in bound
nucleons  (N)  which  differ  from  the  distribution  in  isolated  ones
\cite{ross,nach}. 2) Nuclear  effects due to binding,  the interaction of
nuclear fragments and the like \cite{btom,ak,dun}.   It  is  the  second
alternative we shall pursue below.

The complexity of a relativistic treatment of the EMC effect provides the
standard excuse  for studying non-relativistic (NR)  models.  While being
aware of the many features which have no NR analog, a solution of a model
may  furnish selective  insight.  As  an example  we mention  a proof  of
asymptotic   freedom   of  non-relativistically   confined   constituents
\cite{gr1}.

Most  treatments of  the EMC  effects have  used the  Plane Wave  Impulse
Approximation (PWIA)  which leads to some  form of a convolution  for the
structure function  of the composite nucleus
\cite{atw,rob,jaf2,liu,ji,mul} and
one is inclined  to do the same  in the NR approach.   We shall formulate
instead a  cluster model  with, as central  property, separate  quark and
nucleonic degrees of freedom.  By  construction this assumption enables a
clean analysis of  the nuclear aspects of the EMC  effect.  No additional
approximations are required  to prove that the structure  function of the
fully  interacting  composite  system   is  a  convolution  of  structure
functions for a  nucleon and a nucleus composed  of, respectively, quarks
and point-nucleons. (Section II).  Using the approach by Gersch-Rodriguez
and Smith (GRS) \cite{grs}, we consider the special case of high momentum
transfer.   It permits  a  separation  of the  structure  function in  an
asymptotic   part   and   additional  Final   State   Interaction   (FSI)
contributions,  ordered  by increasing  powers  of  the inverse  momentum
transfer.  In Section  III we consider the above NR  response in the PWIA
for the  nuclear part of the  total response.  In
that treatment we shall be led  to a precise formulation of the off-shell
nature  of  the  nucleonic  component  of  the  total  response.   As  an
alternative to  the above we  shall apply  the West approximation  to the
nuclear  part of  the total  response.  In  Section IV  we first  express
structure  function in  terms of  non-relativistic light-cone  variables.
Then  independent  of the  underlying  cluster  model we  conjecture  the
extension of the convolution formula into the relativistic regime.

Our final expression for the total nuclear structure function contains in
principle  the  complete effect  of  nuclear  dynamics  and serves  as  a
starting-point for the study of  approximations.  We thus discuss for the
nuclear  part  a  relativistic  version  of the PWIA  and a modification,
which includes some part of the FSI effects, neglected in the  PWIA.

\section{Non-relativistic cluster model for the response of a composite
 nucleus.}

We consider a  NR model for inclusive scattering of  a weakly interacting
probe from a  target, transferring to the target momentum  $q$ and energy
$\nu$.  Factoring out from the cross section for the process the same for
the scattering of the probe from a constituent, one is left with, what is
defined as  the response or  structure function  of the target.   For the
latter we  take a nucleus  of $A$ nucleons,  each composed of  3 spinless
$'$quarks$'$ of equal  charge.  With nucleon excitation  energies much in
excess of  the same  for nuclei,  one is led  to a  cluster model  with 3
quarks per nucleon with neglect of Pauli exchange between quarks in
different nucleons.
If the forces  are confining, this is tantamount to a
cluster model with  3-quark bags.

Quarks are thus assumed to move
in a confining potential $V(r_{\ell,i})$,
where ${\mbox  {\boldmath$r$}}_{\ell,i}$ are coordinates of quark $\ell$
relative to ${\mbox {\boldmath $R$}}_i$, the
center-of-mass coordinate of
nucleon $i$ with respect to the center-of-mass of the nucleus.  For those
quark coordinates $\sum_l {\mbox  {\boldmath $r$}}_{l,i}=0$. The nuclear
potential we choose to be of the form
$U=\sum_{j<i}U_{ij}(R_{ij})$, where
$R_{ij}=|{\mbox {\boldmath $R$}}_i-{\mbox {\boldmath  $R$}}_j|$.   The
above implies that the nuclear potential  does not depend on the relative
coordinates of the  quarks {\mbox{\boldmath $r$}'} and likewise, that the
quark   confining  potential   does  not   depend  on  {\boldmath  $R$'}.
Differently stated, we  select Hamiltonians which act  in separate spaces
of quark and nucleon coordinates
\begin{eqnarray}
H=\sum_{i=1}^A \left(\sum_{\ell =1}^3
\left(-\frac{1}{2m}\nabla^2_{\ell ,i}+ V(r_{\ell ,i})\right)
-\frac{1}{2M} \nabla^2_{i}+\sum_{j<i}U(R_{ij})\right),
\label{a1}
\end{eqnarray}
with $m$  and $M$ the  quark and nucleon mass.   As a consequence  of the
above assumption,  states $|\Psi\rangle$  factorize in internal nucleon
and nuclear states $|n\rangle,|N\rangle$ with respective
energies $e_n,E_N$.

For large $q$ the incoherent part dominates the structure function and
its contribution per nucleon is
\begin{eqnarray}
W_{\alpha/{\rm A}}(q,\nu )
=3\sum_n\sum_N {\hspace {-0.5cm}{\int }}\,\,|\langle 0,0|
\exp \left[ i\mbox{\boldmath $q$} ({\mbox{\boldmath ${r}_{11}$}} +
\mbox{\boldmath ${R}_1$})\right] |n,N \rangle|^2\delta
(\nu -E_{N0}-e_{n0}-q^2/2M_{\rm A}),
\label{a2}
\end{eqnarray}
where $E_{N0}=E_N-E_0;\, e_{n0}=e_n-e_0$. The factor in front is
the number of quarks per nucleon. Notice that the spectrum  of confined
quarks is always discrete, while nuclear states $|N\rangle$ may
belong to either discrete or to continuum parts of a spectrum.

In order to evaluate the total structure function
$W_{\alpha/{\rm A}}$ for the composite system, we consider first
the structure function of an isolated nucleon.
For use  below, we also treat the general case of
a  nucleon in  motion with initial momentum ${\mbox{\boldmath  $  P$}}$
and  energy ${\cal E}$. The latter may be
on shell, i.e. ${\cal E}^{on}={\cal E}_{\mbox{\boldmath  $  P$}}=
{\mbox{\boldmath  $  P$}}^2/2M$, but also arbitrary off-shell:
${\cal E}={\cal E}^{off}\ne {\mbox{\boldmath  $  P$}}^2/2M$. The
recoil momentum is ${\mbox{\boldmath $ P$}}+{\mbox{\boldmath $ q$}}$, and
we shall need only the case of on-shell recoil energy,
${\cal E}_{{\mbox{\boldmath $P+q$}}}$. The following expression
\begin{eqnarray}
W_{\alpha/{\rm N}}
({\mbox{\boldmath $ P$}},{\cal E};q,\nu )
=3\sum_{n}\left |\langle 0|\exp\left (i
{\mbox {\boldmath $q r$}}_{11}\right )|n\rangle\right |^2
\delta \bigg (\nu -e_{n0}+ {\cal E}-
{\cal E}_{ {\mbox{\boldmath $ P+q$}}}\bigg )
\label{a3}
\end{eqnarray}
is thus the response for a moving, generally off-shell
nucleon. As Eq. (\ref{a3}) shows,
it is as follows related to the response of an on-shell nucleon at rest
$W_{\alpha/{\rm N}}(q,\nu)\equiv W_{\alpha/{\rm N}}(0,0;q,\nu)$:
\begin{eqnarray}
W_{\alpha/{\rm N}}({\mbox{\boldmath $ P$}},{\cal E};q,\nu)=
W_{\alpha/{\rm N}}(q,\nu+{\cal E}-
{\cal E}_{ {\mbox{\boldmath $P+q$}}}
+{\cal E}_{{\mbox{\boldmath $ q$}}})
\label{a4}
\end{eqnarray}
Next we need the response per nucleon of a nucleus of $A$ point-like
nucleons (in short the nuclear response)
\begin{eqnarray}
W_{{\rm N/A}}(q,\nu )=\sum_{N} {\hspace {-0.5cm}{\int }}\,\,
\left |\langle 0|\exp\left
(i{\mbox {\boldmath $q R$}_1}\right )|N\rangle
\right |^2\delta (\nu-E_{N0}-q^2/2M_{\rm A})
\label{a5}
\end{eqnarray}
{}From Eqs. (\ref{a2}), (\ref{a3}) and (\ref{a5}) one  easily finds for
the  response of the
system under consideration in terms of the  same for a free nucleon and a
nucleus of point  nucleons
\begin{eqnarray}
W_{\alpha/{\rm A}}(q,\nu)
=\int^{\nu}_{q^2/2M_A}W_{{\rm N/A}}(q,\nu')
W_{\alpha/{\rm N}}(q,\nu-\nu'+q^2/2M) d\nu'
\label{a6}
\end{eqnarray}
Eq. (\ref{a6}) is an exact expression  for the desired response  in the
cluster model and is a convolution  of  $W_{\alpha/{\rm N}}$  and
$W_{{\rm N/A}}$:   Nothing but  the  assumption  on separability  of
spaces went in the derivation of the
convolution  (\ref{a6}). Notice that the latter may also be written as
\begin{eqnarray}
W_{\alpha/{\rm A}}(q,\nu)=3\sum_{n}\left |\langle 0|\exp\left
(i{\mbox {\boldmath $q r$}}_{11}\right |
n\rangle\right |^2 W_{{\rm N/A}}(q,\nu-e_{n0}-q^2/2M)
\label{a7}
\end{eqnarray}
Since the energy argument in $W_{{\rm N/A}}$ above is the same
as in the expression
(\ref{a3})  for the  structure function  of a free  on-shell
nucleon at rest  (${\mbox {\boldmath $P$}}=0$),
Eq. (\ref{a7}) shows that embedding of a nucleon in a
nucleus broadens the $\delta$-function peaks in (\ref{a3}),
corresponding to  excitations of the nucleon.

In the framework of our cluster model, which
assumes separability of nucleon and quark interactions,
the response $W_{{\rm N/A}}(q,\nu )$
is an observable quantity. By definition it is that part of the total
response where nucleons are detected without associated particle
production, i.e. in  their ground state $n$=0. Thus
\begin{eqnarray}
W_{{\rm N/A}}(q,\nu)\rightarrow
\frac{\tilde W_{\alpha/{\rm A}}(q,\nu)}{{\cal F}_{\rm N}(q)},
\label{a8}
\end{eqnarray}
where ${\cal  F}_{\rm N}(q)=3\langle
0|\exp\left (i{\mbox  {\boldmath $qr$}}\right)|0 \rangle$ is the elastic
form factor of the nucleon  and $\tilde W_{\alpha/{\rm A}}(q,\nu)$ is the
response  spelled out  above.   In  principle
$\tilde W_{\alpha/{\rm A}}(q,\nu)$ can  be extracted  from
A($e,e')$X$,\,$   A($e,e'$N)X$\;$  A($e,e'$NN)X$\; ,
\dots$   reactions  in
regions where in final states  X
particle production is kinematically forbidden.
$\tilde  W_{\alpha/{\rm  A}}$ is the corresponding  response
integrated over all nucleon momenta.

We now address the high-$q$ behavior of responses and shall use the
Gersch-Rodriguez-Smith (GRS) representation for the response \cite{grs}.
There one replaces
the energy transfer parameter $\nu$ by
the GRS-West variable \cite{grs,wes}
\begin{eqnarray}
y=-\frac{q}{2}+\frac{m\nu}{q}
\label{a9}
\end{eqnarray}
and defines the reduced response
\begin{eqnarray}
\phi(q,y)=(q/m)W(q,\nu)
\label{a10}
\end{eqnarray}
In both (\ref{a9}) and (\ref{a10}), $m$ is always the mass of
the struck particle which absorbs the transferred ($q,\nu)$. As
above we consider  only the dominant
incoherent part  of the NR reduced response. As has been shown  by GRS
\cite{grs} the reduced response permits the expansion
\begin{equation}
\phi (q,y)=F_0(y)+(m/q)F_1(y)+(m/q)^2F_2(y)+\cdots,
\label{a11}
\end{equation}
where the coefficients $F_i$ may  be calculated as function of the
interaction $V$ between the constituents. In particular one has for the
asymptotic limit
\begin{eqnarray}
F_0(y)=\lim_{q \rightarrow \infty}
\phi(q,y)=(q/m) \int n(p)
\delta (\nu +{\epsilon}_{\mbox{\boldmath $ p$}}-
{\epsilon}_{\mbox{\boldmath $p+q$}})d{\mbox{\boldmath $p$}}
=\int n(p)\delta (y-p_z)d{\mbox{\boldmath $p$}},
\label{a12}
\end{eqnarray}
with $n(p)$,
the single constituent momentum  distribution. Eq. (\ref{a12}) shows
that the scaling variable $y$ is the minimal value of the momentum of
the  struck constituent  in  the direction  of ${\mbox  {\boldmath$q$}}$,
when the constituent is on its energy shell before, as well as after the
absorption of the virtual photon.

Note that in the asymptotic limit (\ref{a12})
there is no trace of the interaction $V$ in the single particle energies
${\epsilon}_{\mbox{\boldmath  $p$}}={\mbox{\boldmath $p$}}^2/2m$. This
holds, whether $V$ is regular or singular, corresponding to confining
forces \cite{gr1}. However, that interaction is  implicit in the exact
momentum distribution in (\ref{a12}).

It is now a simple matter to rewrite the convolution (\ref{a6}) in terms
of reduced  responses
\begin{eqnarray}
\phi_{\alpha/{\rm N}}(q,y)=3\sum_{n}\left |\langle 0|\exp\left (i
{\mbox {\boldmath $q r'_{11}$}}\right )|n\rangle
\right |^2\delta\left (y+\frac{M-m}{2M}q-\frac{m e_{n0}}{q}\right )
\label{a13}
\end{eqnarray}
and
\begin{eqnarray}
\phi_{{\rm N/A}}(q,Y)=\sum_N {\hspace {-0.5cm}}{\int }\,\,
\left |\langle 0|\exp\left( i{\mbox {\boldmath $q R'_1$}}\right)|N\rangle
\right |^2 \delta \left(Y+\frac{M_A-M}{2M_A}q-\frac{ME_{N0}}{q} \right )
\label{a14}
\end{eqnarray}
Notice that the appropriate GRS-West scaling
variable in Eq. (\ref{a14}) is
\begin{eqnarray}
Y= -\frac{q}{2}+\frac{M\nu}{q}=\frac{M}{m}\left (y+\frac{M-m}{2M}q\right)
\label{a15}
\end{eqnarray}
One readily checks
\begin{eqnarray}
\phi_{\alpha/{\rm A}} (q,y)=\int_{Y_{min}}^{Y_{max}}
\phi_{{\rm N/A}}(q,Y)\phi_{\alpha/{\rm N}}
\left(q,y-\frac{m}{M}Y\right)dY,
\label{a16}
\end{eqnarray}
where $Y_{min}=-q/2(1-M/M_A)$ and
$Y_{max}=(M/m)y+(q/2)(M/m-1)$.

Next we consider for fixed $y$  asymptotic limits like  Eq. (\ref{a12})
for all responses involved in (\ref{a16})
\begin{mathletters}
\label{a17}
\begin{eqnarray}
\lim_{q\rightarrow\infty}
\phi_{\alpha/{\rm N}}(q,y)
&=&\;(q/m) \int n(p')
\delta (\nu +\epsilon_{\mbox{\boldmath $ p'$}}-
\epsilon_{\mbox{\boldmath $p'+q$}})d{\mbox{\boldmath $p'$}}
=\int n(p')\delta (y-p'_z)d{\mbox{\boldmath $p'$}}
\label{a17a}\\
\lim_{q\rightarrow\infty}
\phi_{{\rm N/A}}(q,Y)
&=&\;(q/M) \int {\cal N}(P)\delta (\nu +
{\cal E}_{\mbox{\boldmath $ P$}}-{\cal E}_{\mbox{\boldmath $P+q$}})
d{\mbox{\boldmath $P$}}
=\int {\cal N}(P)\delta (Y-P_z)d{\mbox{\boldmath $P$}},
\label{a17b}
\end{eqnarray}
\end{mathletters}
where $n(p')$  and ${\cal N}(P)$ are the single  quark, respectively  the
single nucleon momentum distributions. Applying Eqs. (17) to (\ref{a16})
one obtains
\begin{eqnarray}
\lim_{q\rightarrow\infty} \phi_{\alpha/{\rm A}}(q,y)=
\int\int {\cal N}(P)n(p')\delta\bigg (y-p'_z-(m/M)P_z\bigg )
d{\mbox{\boldmath $P$}}d{\mbox{\boldmath $p'$}}
=\int {\tilde n}(p)\delta (y- p_z )d{\mbox{\boldmath $p$}},
\label{a18}
\end{eqnarray}
with the momentum distribution
\begin{eqnarray}
\tilde n(p)=\int {\cal N}(P)n\bigg(
{\mbox{\boldmath $p$}}-(m/M){\mbox{\boldmath $P$}}\bigg)
d{\mbox{\boldmath $P$}},
\label{a19}
\end{eqnarray}
generated by convoluting the distributions ${\cal N}$ and $n$.
With  ${\mbox{\boldmath  $P$}}$  and
${\mbox{\boldmath $p$}}$ the momentum of
the nucleon and the struck quark,
${\mbox{\boldmath $ p$}}-(m/M){\mbox{\boldmath $P$}}$
is  the relative  momentum  of  that  quark  with respect  to  the
center-of-mass of the nucleon. (See Fig. 1  for the kinematics of
the  three situations considered: Eqs. (17),  and (\ref{a18})).
Eq. (\ref{a18})  manifests that also quarks, confined
in nucleons and in turn bound in nuclei, are asymptotically free.

We note from  (\ref{a18}) that the location  of the Quasi-Elastic
Peak (QEP) in the asymptotic response  is the same as for a nucleus of
point-nucleons,
namely at $y=0$.  Eq. (\ref{a19}) shows that the composite nature of the
target, affects only the width given by $n(p)$: A $q$-dependent shift in
the position of  the above $'$asymptotic$'$ QEP is generated by
higher order FSI contributions in (\ref{a11}) and is therefore
not manifestly related to nuclear binding.

\section{The response of a composite target in the PWIA.}

Virtually all treatments of the EMC effect are based on the PWIA
\cite{btom,ross,nach,ak,dun,atw,rob,jaf2,liu,ji,mul}
which we now briefly review.  In that approximation
for the nuclear part  of the Hamiltonian (\ref{a1}) one neglects
the interaction between a selected nucleon $'1'$ and the remaining
($A$-1)-particle core, i.e.
$\sum_{j<i}U(R_{ij})\rightarrow \sum_{2\le j<i}U(R_{ij})$. Excited
states in Eq. (\ref{a5}) then become
\begin{eqnarray}
|N\rangle \rightarrow |N\rangle^{PWIA}=
|(A-1)_{\lambda,-{\mbox{\boldmath $ P$}} };{\mbox{\boldmath $P$}}\rangle,
\label{a20}
\end{eqnarray}
where $|(A-1)_{\lambda,-{\mbox{\boldmath $ P$}} }{\mbox{\boldmath $ P$}}
\rangle$  denotes an excited state of a moving core
with internal and kinetic energies ${\cal E}^C_{\lambda},
{\cal E}^C_{{\mbox{\boldmath $ P$}}}$,  and a free nucleon with momentum
${\mbox{\boldmath $P$}}$ and energy ${\cal E}_{{\mbox{\boldmath $ P$}}}$.
Substitution of (\ref{a20}) into (\ref{a5}) and the replacements
$\displaystyle \sum_N{\hspace {-0.5cm}{\int }}\rightarrow
\displaystyle \sum_{\lambda} {\hspace {-0.5cm}{\int }}
\int d{\mbox{\boldmath $ P$}}\to\int dE d{\mbox{\boldmath $ P$}}$ lead to
\begin{mathletters}
\label{a21}
\begin{eqnarray}
W_{{\rm N/A}}^{PWIA}(q,\nu)&=&\displaystyle
\sum_{\lambda}{\hspace {-0.5cm}{\int }}\int
|{\psi}_{\lambda}({\mbox{\boldmath $ P$}} )|^2
\delta(\nu-\Delta_{\lambda}-{\cal E}^C_{{\mbox{\boldmath $ P$}} }-
{\cal E}_{ {\mbox{\boldmath $P$}} +{\mbox{\boldmath $ q$}} })
d{\mbox{\boldmath $ P$}}
\label{a21a}\\
&=&\int dE\int d{\mbox{\boldmath $ P$}}
S_{{\rm N/A}}({\mbox{\boldmath $ P$}},E)
\delta (\nu -E-{\cal E}^C_{{\mbox{\boldmath $ P$}}}
-{\cal E}_{{\mbox{\boldmath $ P+q$}}})
\label{a21b}\\
&=&\int d{\mbox{\boldmath $ P$}} S_{{\rm N/A}}
({\mbox{\boldmath $ P$}},\nu -{\cal E}^C_{{\mbox{\boldmath $ P$}}}
-{\cal E}_{{\mbox{\boldmath $ P+q$}}})
\label{a21c}
\end{eqnarray}
\end{mathletters}
where ${\psi}_{\lambda}({\mbox{\boldmath $ P$}})=\langle(A-1)_
{\lambda,-{\mbox{\boldmath $ P$}} };{\mbox{\boldmath $P$}}|0\rangle$,
is the overlap of the ground state of the target at rest and
an excited core state $\lambda$ with a free nucleon. The function
\begin{eqnarray}
S_{{\rm N/A}}({\mbox{\boldmath $ P$}},E)
&=&\displaystyle \sum_{\lambda}{\hspace {-0.5cm}{\int }}
|{\psi}_{\lambda}({\mbox{\boldmath $ P$}} )|^2
\delta(E-\Delta_{\lambda})
\label{a22}
\end{eqnarray}
above is the single-particle spectral function in terms of the
separation  energy $\Delta_{\lambda}$ of a nucleon, when removed from
the target ground state, and leaving the core in an excited state
$\lambda$. One  notices that the integrand in (\ref {a21c})
is the structure function for the semi-inclusive A$(e,e'p)$X reaction
in the PWIA, (Fig. 2), i.e. the  spectral function in (\ref{a22}) where
the energy argument is replaced by the missing energy
$\nu-{\cal E}^C_{{\mbox{\boldmath $ P$}} }-
{\cal E}_{ {\mbox{\boldmath $P$}} +{\mbox{\boldmath $ q$}}}$,
as required by energy conservation \cite{moug}.
The integral over the nucleon momentum $\mbox{\boldmath{$P$}}$ leads
to the structure function for the totally  inclusive A$(e,e')$X process
in that approximation, with no produced particles in final states
$X$ (cf. Eq. (8)).

A pre-requisite for the application of the PWIA to the nuclear response
is an explicit Hamiltonian. However, in view of the singular, confining
inter-quark forces, and in line with many other authors, we do not
detail the $'$quark$'$ part of $H$, or any approximation to it.
Subsequent application of the PWIA to the remaining part of $H$
requires, strictly speaking, that part to be
free  of quark degrees  of freedom, which almost guides one to the
cluster  Hamiltonian   (\ref{a1}). When in the following we shall refer
to PWIA, we  have in mind application of that approximation exclusively
to the nuclear component in a convolution, thus
\begin{eqnarray}
W_{\alpha/{\rm A}}^{PWIA}(q,\nu)\equiv
\int^{\nu}_{q^2/2M_A}W_{{\rm N/A}}^{PWIA}(q,\nu')
W_{\alpha/{\rm N}}(q,\nu-\nu'+q^2/2M) d\nu'
\label{a99}
\end{eqnarray}
Substitution of (\ref{a21a}) into Eq. (\ref{a99}) yields (cf. Fig. 3)
\begin{eqnarray}
W_{\alpha/A}^{PWIA}(q,\nu)
&=&\displaystyle
\sum_{\lambda}{\hspace {-0.5cm}{\int }}
\int d{\mbox{\boldmath $ P$}}
|{\psi}_{\lambda}({\mbox{\boldmath $ P$}} )|^2
W_{\alpha/{\rm N}}(q,\nu-
\Delta_{\lambda}-{\cal E}^C_{{\mbox{\boldmath $ P$}} }
-{\cal E}_{ {\mbox{\boldmath $ P$}}+{\mbox{\boldmath $ q$}}}+
{\cal E}_{{\mbox{\boldmath $ q$}}})
\nonumber\\
&=& \int_{E_m}^{E_M} dE
\int d{\mbox{\boldmath $ P$}}
S_{N/A}({\mbox{\boldmath $ P$}},E)
W_{\alpha/{\rm N}}({\mbox{\boldmath $ P$}},
{\cal E}^{off}({\mbox{\boldmath $ P$}},E);q,\nu),
\label{a23}
\end{eqnarray}
with  upper and lower limits $E_m, E_M$, resulting from the integrates
over the $\delta$-function in (\ref{a21b}).
Comparison with Eqs. (\ref{a4}) shows that the structure function
$W_{\alpha/N}$ relates to a nucleon target with initial
momentum ${\mbox{\boldmath $ P$}}$  and  off-shell energy
\begin{eqnarray}
{\cal E}={\cal E}^{off}({\mbox{\boldmath $ P$}},E)=
-E-{\cal E}^C_{\mbox{\boldmath $ P$}}=
-\Delta_{\lambda}-{\cal E}^C_{{\mbox{\boldmath $ P$}} }
\label{a24}
\end{eqnarray}
Indeed, the  PWIA prescription, displayed in Fig. 2  assigns the
nuclear target and the spectator core  (but not the struck particle)
to be on their respective energy shells:  Energy conservation then
prescribes the off-shell energy (\ref{a24}).

As in the previous Section, we replace the energy transfer
$\nu$ by the scaling variable $y$, Eq. (\ref{a9}), and
obtain  for the nuclear reduced response (\ref{a21b})
\begin{eqnarray}
\phi^{PWIA}_{{\rm N}/{\rm A}}(q,Y)=
\int dE\int d{\mbox{\boldmath $ P$}}
S_{{\rm N/A}}({\mbox{\boldmath $ P$}},E)
\delta \left ( Y-P_z-\frac{M}{q}(E+{\cal E}^C_{{\mbox{\boldmath $ P$}}}+
{\cal E}_{{\mbox{\boldmath $ P$}}})\right )
\label{a27}
\end{eqnarray}
The total reduced response corresponding to (\ref{a99}) thus becomes
\begin{eqnarray}
\phi_{\alpha/{\rm A}}^{PWIA}(q,y)
=\int dY\phi^{PWIA}_{{\rm N}/{\rm A}}(q,Y)
\phi_{\alpha/{\rm N}}(q,y-\frac{m}{M}Y)
\label{a28a}
\end{eqnarray}
Substitution of (\ref{a27}) in (\ref{a28a}) and subsequent integration
over $Y$ yields
\begin{eqnarray}
\phi_{\alpha/{\rm A}}^{PWIA}(q,y)
=\int_{E_m}^{E_M} dE\int d{\mbox{\boldmath $ P$}}
S_{\rm {N/A}}({\mbox{\boldmath $ P$}},E)
\phi_{\alpha/{\rm N}}\bigg ({\mbox{\boldmath $ P$}},
{\cal E}^{off}({\mbox{\boldmath $ P$}},E);q,y\bigg )
\label{a25}
\end{eqnarray}
$\phi_{\alpha/{\rm N}}$ above is the response of a moving
nucleon  with off-shell energy  ${\cal E}^{off}$, Eq. (\ref{a24}). It
can be related to the same at rest using the equivalent of Eq. (\ref{a4})
in $y$
\begin{mathletters}
\label{a26}
\begin{eqnarray}
\phi_{\alpha/{\rm N}}({\mbox{\boldmath $ P$}}, {\cal E}^{off};q,y)
&=&\phi_{\alpha/{\rm N}}\bigg (q,y-\frac{m}{M}P_z+\frac{m}{q}
({\cal E}^{off}-
{\cal E}_{{\mbox{\boldmath $ P$}}})\bigg )
\label{a26a}\\
\phi_{\alpha/{\rm N}}({\mbox{\boldmath $ P$}}, {\cal E}^{on};q,y)
&=&\phi_{\alpha/{\rm N}}\left (q,y-\frac{m}{M}P_z\right ),
\label{a26b}
\end{eqnarray}
\end{mathletters}
Along (\ref{a25}), an  alternative expression may be derived from
Eqs. (\ref{a27}) and (\ref{a28a}), integrating now over $E$
\begin{eqnarray}
\phi_{\alpha/{\rm A}}^{PWIA}(q,y)
=\frac {q}{M} \int dY\int d{\mbox{\boldmath $ P$}}
S_{{\rm N/A}}\left ({\mbox{\boldmath $ P$}},\frac{q}{M}(Y-P_z)-
({\cal E}^C_{{\mbox{\boldmath $ P$}}}+
{\cal E}_{{\mbox{\boldmath $ P$}}})\right )
\phi_{\alpha/{\rm N}}(q,y-\frac{m}{M}Y)
\label{a28b}
\end{eqnarray}
Compare now Eqs. (\ref{a25}) and (\ref{a28b}).  The former contains  a
${\mbox{\boldmath $P$}}$-dependent response of an {\it{off}}-shell,
moving nucleon. In  contradistinction, $\phi_{\alpha/{\rm N}}$ in
(\ref{a28b}) is the ${\mbox{\boldmath $ P$}}$-independent response of an
{\it{on}}-shell nucleon, which enables  formal integration
over the momentum ${\mbox{\boldmath  $P$}}$ in the nuclear response, Eq.
(\ref{a28b}). In spite of their  apparent dissimilarity,
Eqs.  (\ref{a25}), (\ref{a28b}) are identical  expressions for  the total
response in the PWIA, obtained by performing  the $Y$, respectively the
$E$ integrations in Eqs. (\ref{a27}), (\ref{a28a}).

We conclude this section by mentioning the alternative West
approximation for the  reduced  nuclear response  \cite{wes}, where
out of the GRS  series,   Eq.  (\ref{a11}), only the first term
$\phi^W_{{\rm  N}/{\rm A}}(y)=  F_0(y)=\lim_{q\rightarrow \infty}
\phi_{{\rm N}/{\rm  A}}(q,y)$,  (\ref{a17b}) is retained. It leads to
\begin{eqnarray}
\phi^W_{\alpha/{\rm A}}(q,y)=\int dY \phi^W_{{\rm N/A}}(Y)
\phi_{\alpha/{\rm N}}(q,y-\frac{m}{M}Y)= \int d{\mbox{\boldmath $ P$}}
{\cal N}({\mbox{\boldmath $ P$}})\phi_{\alpha/{\rm  N}}
({\mbox{\boldmath  $P$}}, {\cal E}^{on};q,\nu),\label{a29}
\end{eqnarray}
i.e.  the nucleon
response averaged  over the nuclear Fermi  momentum-distribution.  Notice
that,  contrary  to  the  PWIA expression  Eq. (\ref{a25}),  the  above
approximation causes the energy of the  struck nucleon to be shifted back
to its on-shell value, as in Eq. (\ref{a26b}). For underlying NR
kinematics
$$\lim_{q \rightarrow \infty} \int_{E_m}^{E_M} dE S_{\rm N/A}
(E,{\mbox{\boldmath $ P$}})=\int _0^{\infty} dE S_{\rm N/A}
(E,{\mbox{\boldmath $ P$}})={\cal  N}({\mbox{\boldmath $ P$}}),$$
witn {\cal N}, the single nucleon  momentum distribution.
One  checks from Eqs.  (\ref{a25}), (\ref{a29}) that in that limit
the PWIA and the West approximations for $\phi_{{\rm N}/{\rm A}}$
coincide.

\section{Relativistic extension}

In order to perform a transition of the above results
to  the relativistic case, we
introduce light-cone variables as an intermediary.
Thus for a quark ($'\alpha'$), a nucleon and the nuclear target at rest
with 4-momentum $p, P$ and $P_{\rm A}$ we define  correspondingly
light-cone momenta for each kind of particles, thus
$p_{\pm}=p_0\pm p_z;\, P_{\pm}=P_0\pm P_z; \,P^A_{\pm}=M_A$. In addition
we shall need light-cone fractions\   $p_{-}/P^{\rm A}_{-};\,
P_{-}/P_{-}^{\rm A}$ which we choose to relate to a nucleon, and not
to the  actual nuclear target at rest. Thus
$x_{\alpha/{\rm A}}\equiv p_{-}/M;\, x_{{\rm N}/{\rm A}}\equiv P_{-}/M$.

Consider now the above light-cone  variables in the NR limit. In
accordance with Eqs. (17), $p_z\rightarrow y;\ \, P_z\rightarrow Y$ and
we thus have the following correspondences
\cite{fnote}
\begin{mathletters}
 \label{a30}
\begin{eqnarray}
x_{\alpha/A}&=&\,\frac{p_0-p_z}{M} \,\,\rightarrow \xi_{\alpha/A}=
\frac{m-y}{M}  \\
\label{a30a}
x_{N/A}&=&\frac{P_0-P_z}{M} \,\rightarrow \xi_{N/A}=
\frac{M-Y}{M}
\label{a30b}
\end{eqnarray}
\end{mathletters}
Those define structure functions of on-shell targets at rest
in terms of those NR momentum fractions
\begin{mathletters}
\label{a31}
\begin{eqnarray}
f^{nr}_{\alpha /{\rm N}}(q,\xi )
&\equiv& M\phi_{\alpha/{\rm N}}(q,m-M\xi)
\label{a31a}\\
f^{nr}_{{\rm N/}{\rm A}}(q,\xi )
&\equiv& M\phi_{{\rm N}/{\rm A}}(q,M-M\xi)
\label{a31b}\\
f^{nr}_{\alpha /{\rm A}}(q,\xi )&\equiv&
M\phi_{\alpha/{\rm A}}(q,m-M\xi),
\label{a31c}
\end{eqnarray}
\end{mathletters}
Eq. (\ref{a26a}) enables the extension of the above, to structure
functions of a moving off-shell nucleon
\begin{eqnarray}
f^{nr}_{\alpha /{\rm N}}({\mbox{\boldmath $P$}},{\cal E};q,\xi )
\equiv M\phi_{\alpha/{\rm N}}\left ( q,y -\frac{m}{M}P_z
-\frac{m}{q}\Delta{\cal E}\right )\cong f^{nr}_{\alpha /{\rm N}}
\left (q,\frac{M\xi}{M(1-\Delta{\cal E}/q)-P_z}\right ),
\label{a32}
\end{eqnarray}
where $\Delta{\cal E}={\cal E}_{\mbox{\boldmath $ P$}}-{\cal E}$ is
the off-shell energy shift. In the last step in (\ref{a32}) we have
used the approximation
$$y-\frac{m}{M}P_z-\frac{m}{q}\Delta{\cal E}\cong m-\frac {M(m-y)}{M(1-
\Delta{\cal E}/q))-P_z}$$
Just as  Eqs. (\ref{a6}), (\ref{a16})  express the nuclear response  as a
convolution in  $\nu$ and  a  NR scaling  variable $y$, one obtains
a third representation  of the  total response  in terms  of the
variables  $\xi$. Substitution of (\ref{a30})  into  (\ref{a16}), the use
of the definitions (\ref{a31}) and the
approximation $m-y+(m/M)Y\cong M(\xi_{\alpha/{\rm A}}/
\xi_{{\rm N/A}})$ leads to
\begin{mathletters}
\label{a33}
\begin{eqnarray}
f^{nr}_{\alpha/{\rm A}}(q,\xi_{\alpha/{\rm A}})&=&
\int d\xi_{{\rm N}/{\rm A}}
f^{nr}_{{\rm N}/{\rm A}}\left (q,\xi_{{\rm N}/{\rm A}} \right )
f^{nr}_{\alpha/{\rm N}} \bigg(q,\xi_{\alpha/{\rm A}}-
\frac{m}{M}\xi_{{\rm N}/{\rm A}}+\frac{m}{M}\bigg)
\label{a33a}\\
&\cong& \int d\xi_{{\rm N}/{\rm A}} f^{nr}_{{\rm N}/{\rm A}}
\left(q,\xi_{{\rm N}/{\rm A}}\right ) f^{nr}_{\alpha/{\rm N}} \bigg( q,
\frac{\xi_{\alpha/{\rm A}}}{\xi_{{\rm N}/{\rm A}}} \bigg)
\label{a33b}
\end{eqnarray}
\end{mathletters}

A particular case is the asymptotic limits for
the structure functions in Eq. (\ref{a33})
\begin{mathletters}
\label{a34}
\begin{eqnarray}
f^{nr}_{\alpha /{\rm N}}(\xi )=
\lim_{q\rightarrow\infty}f^{nr}_{\alpha /{\rm N}}(q,\xi )
&=&M\int d{\mbox{\boldmath $ p_{\perp}$}}
n({\mbox{\boldmath $ p_{\perp}$}},m-M\xi)
\label{a34a}\\
f^{nr}_{{\rm N}/{\rm A}}(\xi )=
\lim_{q\rightarrow\infty}f^{nr}_{{\rm N}/{\rm A}}(q,\xi )
&=&M\int d{\mbox{\boldmath $ P_{\perp}$}}
{\cal N}({\mbox{\boldmath $ P_{\perp}$}},M-M\xi),
\label{a34b}
\end{eqnarray}
\end{mathletters}
i.e. probabilities to find, respectively a quark inside the nucleon
and a nucleon inside the nucleus, with  light-cone momentum fraction
$\xi$. Eq. (\ref{a31c}) then becomes
\begin{eqnarray}
f^{nr}_{\alpha/{\rm A}}(\xi_{\alpha/{\rm A}})
=\lim_{q \rightarrow \infty}
f^{nr}_{\alpha/{\rm A}}(q,\xi_{\alpha/{\rm A}})
=M\int d{\mbox{\boldmath $ p'_{\perp}$}} \tilde {n}
\bigg ({\mbox{\boldmath $ p'_{\perp}$}},m-M\xi_{\alpha/{\rm A}}\bigg ),
\label{a35}
\end{eqnarray}
with $\tilde n$ as in Eqs. (\ref{a18}), (\ref{a19}).

Eqs. (\ref{a33}) are a nearly direct consequence of a strictly
NR cluster model, which is of course radically different from a realistic
rendition of the actual relativistic  dynamics for the
constituents. Notwithstanding, the above result bears clear features of
relativistic expressions for the total nuclear structure function which
appeared in the literature. This observation invites to dissociate the
outcome (\ref{a33}) from the underlying model and to attempt an
extension into the relativistic domain.

We first return
to the relativistic light-cone fractions in Eq. (\ref{a30})
and then conjecture the validity of Eqs. (\ref{a33}),
if NR quantities are replaced by relativistic ones. Thus with
$Q^2=q^2-\nu^2$ and
\begin{eqnarray}
f^{nr}(q,\xi) \to f(Q^2,x)
\label{a36}
\end{eqnarray}
Eq. (\ref{a33b}) becomes
\begin{eqnarray}
f_{\alpha/{\rm A}}(Q^2,x_{\alpha/{\rm A}})=
\int^{M_A/M}_{x_{\alpha/{\rm A}}}dx_{{\rm N}/{\rm A}}
f_{{\rm N}/{\rm A}}(Q^2,x_{{\rm N}/{\rm A}})
f_{\alpha/{\rm N}} \bigg(Q^2,\frac{x_{\alpha/{\rm A}}}
{x_{{\rm N}/{\rm A}}}\bigg)
\label{a37}
\end{eqnarray}
The definitions (\ref{a30}) connect the light-cone momenta
$P_{-},p_{-}$ to the fractions
$x$, and those in turn to the Bjorken
variables. For instance $x_{{\rm N}/{\rm A}}=(P_0-P_z)/M$,
while  in the high $q$-limit
the corresponding Bjorken variable becomes
$x^{Bj}_{{\rm N}/{\rm A}}= Q^2/2M\nu \to (q-\nu)/M$. Energy
conservation, and the fact that the knocked-out nucleon is on its mass
shell (cf. Fig. 2),  fix in the same limit $P_0=
[({\mbox{\boldmath $P+q$}})^2+M^2]^{1/2}-\nu
\rightarrow P_z+q-\nu$.
The two quantities approach therefore the same limit in the large
$q,\nu$ limit: we conjecture that (\ref{a37}) holds with
$x$ interpreted as $x^{Bj}$ for any finite $Q^2$.

Eq. (\ref{a37}) is our main result. It gives the distribution
function of a fully composite nucleus in terms of the same for
point nucleons and for an isolated nucleon. The emphasis is on
the distribution $f_{{\rm N}/{\rm A}}$, which in principle
contains the complete inter-nucleon dynamics.

At this point we focus on the difference in treatment of inter-quark
and inter-nucleon forces. No mention has been made of the former
and we deliberately avoid a difficult, realistic calculation of
$f_{\alpha /{\rm N}}$, the hard portion in the total structure function.
Instead one would like to take it from data.
In principle no such problem exists for a calculation of
$f_{{\rm N/A}}$, the soft nuclear part of the
total structure function, and which may be described using conventional
nuclear physics. We now proceed with a discussion of two approximations
for the nuclear structure function and start with the PWIA.

Assume that, as above, the nuclear states can be described
non-relativistically and the same for derived quantities
$\psi_{\lambda}$ or the spectral function $S_{{\rm N/A}}$,
Eqs. (\ref{a20})-(\ref{a22}). The sole relativistic aspect retained
in this soft component is the kinematics of the knocked-out nucleon.
We thus write for the structure function $W_{{\rm N/A}}(q, \nu )$
in the relativistic PWIA (cf. Fig. 2)
\begin{eqnarray}
W_{{\rm N}/{\rm A}}^{PWIA}(Q^2,\nu )=
\displaystyle
\sum_{\lambda}{\hspace {-0.5cm}{\int }}
\int d^4P\, 2M|{\psi}_{\lambda}({\mbox{\boldmath $ P$}} )|^2
\delta\left [ (P_0+\nu )^2-({\mbox{\boldmath $P+q$}})^2-M^2\right ]
\delta (P_0-P_0^{off}),
\label{a38}
\end{eqnarray}
where
\begin{eqnarray}
P_0^{off}=M_A-[(M^{\lambda}_{A-1})^2
+{\mbox{\boldmath $ P$}}^2]^{1/2}-\nu
\cong M-\Delta_{\lambda}-{\cal E}^C_{\mbox{\boldmath $ P$}}
\label{a39}
\end{eqnarray}
is the off-shell total energy of the struck nucleon.
The corresponding structure function $f_{\rm N/A}$ is obtained
from $W_{\rm N/A}$ by replacing $\nu$ by $Q^2/2Mx_{\rm N/A}$. Thus
using Eqs. (10), (\ref{a31}), (\ref{a33b}) and (\ref{a37})
the total nuclear response in the PWIA (see Fig. 3) becomes
\begin{mathletters}
\label{a40}
\begin{eqnarray}
f^{PWIA}_{\alpha/{\rm A}}(Q^2,x_{\alpha/{\rm A}}) &=&
\int^{M_A/M}_{x_{\alpha/{\rm A}}}dx_{{\rm N}/{\rm A}}
f^{PWIA}_{{\rm N}/{\rm A}}(Q^2,x_{{\rm N}/{\rm A}})
f_{\alpha/{\rm N}} \bigg(Q^2,\frac{x_{\alpha/{\rm A}}}
{x_{{\rm N}/{\rm A}}}\bigg)
\label{a40a}\\
%\end{document}
f_{{\rm N/A}}^{PWIA}(Q^2,x_{{\rm N/A}})&=&
\nu W_{{\rm N}/{\rm A}}^{PWIA}(Q^2,Q^2/2Mx_{{\rm N/A}}),
\label{a40b}
\end{eqnarray}
\end{mathletters}
Consider now Eqs. (\ref{a40})
in the large-$q$ limit $q^2\gg M^2+{\mbox{\boldmath $ P$}}^2$.
It permits to approximate to write in
(\ref{a38}) $[({\mbox{\boldmath $P+q$}})^2+M^2]^{1/2}\rightarrow q+P_z$
while in the large-$q$ limit
$x_{\rm N/A}\rightarrow (q-\nu )/M$. As a result Eq. (\ref{a40b}) becomes
\begin{eqnarray}
f^{PWIA}_{{\rm N}/{\rm A}}(Q^2,x_{{\rm N}/{\rm A}}) \rightarrow
\displaystyle \sum_{\lambda}{\hspace {-0.5cm}{\int }}
\int d{\mbox{\boldmath $ P$}}
|{\psi}_{\lambda}({\mbox{\boldmath $ P$}} )|^2
\delta\bigg (x_{{\rm N}/{\rm A}}-\frac{P^{off}_0-P_z}{M}\bigg )
\label{a41}
\end{eqnarray}
Using the definition (\ref{a22}) of the single nucleon spectral function,
substitution of (\ref{a41}) into (\ref{a40a}) gives on the one hand,
after integration over $x_{{\rm N/A}}$ (cf. \cite{cl})
\begin{eqnarray}
f^{PWIA}_{\alpha/{\rm A}}(Q^2,x_{\alpha /{\rm A}})=
\int^{E_M}_{E_m} dE \int d{\mbox{\boldmath $ P$}}
S_{\rm N/A}({\mbox{\boldmath $P$}},E)f_{\alpha /{\rm N}}\left (
{\mbox{\boldmath $P$}},P_0^{off}({\mbox{\boldmath $P$}},E);
Q^2,x_{\alpha/{\rm A}}\right ),
\label{a42}
\end{eqnarray}
where (cf. Eq. (\ref{a32}))
\begin{eqnarray}
f_{\alpha /{\rm N}}\left (
{\mbox{\boldmath $P$}},P_0^{off}({\mbox{\boldmath $P$}},E);
Q^2,x_{\alpha/{\rm A}}\right )=
f_{\alpha /{\rm N}}\left (
Q^2,\frac{Mx_{\alpha/{\rm A}}}{P_0^{off}-P_z}\right )
\label{a43}
\end{eqnarray}
is the structure function for a moving nucleon with off-shell energy
$P_0^{off}=M-E-{\cal E}^C_{\mbox{\boldmath $P$}}$. On the other hand,
integration over $E$ leads to a  nucleon structure
function with no explicit {\mbox{\boldmath $ P$}} dependence
\begin{mathletters}
\label{a44}
\begin{eqnarray}
f^{PWIA}_{\alpha/{\rm A}}(Q^2,x_{\alpha/{\rm A}})
&=& \int^{M_A/M}_{x_{\alpha/{\rm A}}}dx_{{\rm N}/{\rm A}}
f^{PWIA}_{{\rm N}/{\rm A}}(x_{{\rm N}/{\rm A}})
f_{\alpha/{\rm N}} \bigg(Q^2,\frac{x_{\alpha/{\rm A}}}
{x_{{\rm N}/{\rm A}}}\bigg)
\label{a44a}\\
f^{PWIA}_{\rm N/A}(x_{\rm N/A})
&=&\int d{\mbox{\boldmath $ P$}}S_{\rm N/A}\left (
{\mbox{\boldmath $ P$}}, M(1-x_{\rm N/A})
-P_z-{\cal E}^C_{\mbox{\boldmath $ P$}})\right )
\label{a44b}
\end{eqnarray}
\end{mathletters}
The expressions (\ref{a42}) and (\ref{a44a}) appear to differ
in the explicit
{\mbox{\boldmath $ P$}}-dependence  of  the   off-shell,  respectively
{\mbox{\boldmath $ P$}}-independence of the on-shell  nucleon structure
function $f_{\alpha/{\rm N}}$. Those are actually  identical as
are the NR expressions (\ref{a25}), (\ref{a28b}).

The PWIA is only one particular approximation for the nuclear part of
the total response  and parallel
with the above NR development, we now discuss a possible relativistic
version  of the alternative West approach (cf. Eq. (\ref{a29})).
In the NR case the GRS theory provides the series (11), and in
particular the asymptotic limit $F_0(y)$ in terms of the single nucleon
momentum distribution ${\cal N}(P)$ and a scaling variable $y$. The
latter we recall results in the NR case if, before and after absorption
of the virtual photon, the nucleon is on its (energy-)shell.
Regarding a relativistic generalization of $y$, the use of relativistic
kinematics under similar on-mass shell conditions does not
produce an acceptable expression. One needs in fact some theoretical
framework for a proper definition. An example is the Bethe-Salpeter
equation, where  to lowest order as in the relativistic PWIA,
the spectator is on its mass-shell. If
the above nucleon before and after absorption of the virtual photon, is
in equal measure kept  off its mass-shell, i.e.
$[ P_0^2-{\mbox{\boldmath $P$}}^2 ] =
[ (P_0+\nu )^2-({\mbox{\boldmath $P+q$}})^2 ]$.
the above  suffices for the definition of a proper relativistic analog
of $y$ \cite{gurv}. One obtains
\begin{eqnarray}
W_{{\rm N}/{\rm A}}^W(Q^2,\nu )&=&
\displaystyle 2M\sum_{\lambda}{\hspace {-0.5cm}{\int }}
\int d^4P\, |{\psi}_{\lambda}({\mbox{\boldmath $ P$}} )|^2
\delta\left [ P_0^2-{\mbox{\boldmath $P$}}^2-(P_0+\nu )^2+
({\mbox{\boldmath $P+q$}})^2\right ]  \delta (P_0-P_0^{off})
\label{a45}
\end{eqnarray}
which leads to the total response
\begin{eqnarray}
f^{W}_{\alpha/{\rm A}}(Q^2,x_{\alpha/{\rm A}})
&=& \int^{M_A/M}_{x_{\alpha/{\rm A}}}dx_{{\rm N}/{\rm A}}
f^{W}_{{\rm N}/{\rm A}}(Q^2,x_{{\rm N/A}})
f_{\alpha/{\rm N}} \bigg(Q^2,\frac{x_{\alpha/{\rm A}}}
{x_{{\rm N}/{\rm A}}}\bigg)
\nonumber\\
f^W_{\rm N/A}(Q^2,x_{\rm N/A})&=&\nu W^W_{\rm N/A}(Q^2,\nu)\nonumber\\
&=&\int d{\mbox{\boldmath $ P$}}S_{\rm N/A}\left (
{\mbox{\boldmath $ P$}},M(1-x_{\rm N/A})-
\left (1+\frac {4M^2x_{\rm N/A}^2}{Q^2}
\right)^{1/2}P_z -{\cal E}^C_{\mbox{\boldmath $ P$}}\right ),
\label{a46}
\end{eqnarray}
with $P_0^{off}$ again as in (\ref{a39}).  The above approximation and
the relativistic PWIA, Eqs. (\ref{a44b}) differ
for finite $Q^2$, but tend to the result (\ref{a44b}) for
$Q^2\rightarrow\infty$, $x_{\alpha /{\rm A}}$ =const. That limit appears
not directly related to a Fermi averaged nucleon distribution function,
as is the NR analog (\ref{a29}). Other treatments of FSI in the nuclear
response will be mentioned in the Discussion.

\section{Summary and Discussion}

We have studied  above inclusive scattering from a  nucleus with nucleons
composed of spinless quarks.  We assumed  a model with triplets of quarks
per  nucleon,  which do  not  interact  with  quarks in  other  nucleons.
Likewise,  the  interactions  between  nucleons  is  independent  of  the
positions of the constituents.  The underlying assumptions are those of a
cluster model  with separate  quark and nucleon  dynamics.  If  the quark
interactions confine, the specific cluster  model is one with exclusively
3-quark bags.

The  above  cluster  model  is naturally  realized  for  non-relativistic
dynamics  and implies  then  separated spaces  for  nucleons and  quarks.
Without the need to specify the involved interactions, the model suffices
for a proof that the nuclear  structure function for the composite system
is a convolution of those functions  for a nucleon and a nucleus composed
of,  respectively,  quarks  and fully  interacting  point-nucleons.   The
result  generalizes the  same  used in  the  PWIA and  is  valid for  any
nucleon-nucleon (or  nucleon-core) interaction beyond the  PWIA.  As such
the model is an ideal tool to  study nuclear facets of the EMC effect, in
spite of the  impossibility to produce excited states of  the nucleons by
the purely nuclear part of the cluster Hamiltonian.

Dependent  on the  choice of  the external  variable in  addition to  the
momentum  transfer $q$  (e.g.  the  energy transfer  $\nu$  or a  scaling
variable $y$) appropriate expressions can  be given for the participating
structure  functions.   We emphasized  in  particular  the NR  limits  of
light-cone variables.  The form of the resulting total structure function
has  then been  disconnected from  the  underlying cluster  model and  by
proper change to  the relativistic light-cone variables,  assumed to hold
in the  relativistic regime.  A  novel feature  is a proper  inclusion in
principle, of the full nucleon-nucleon interaction in the nuclear part of
the Hamiltonian.

Starting from that result, we then studied two approximations.  The first
is the  PWIA, which has been  the standard tool for  the investigation of
the nuclear response, and where the interaction between a nucleon and the
remaining  core  is   neglected.   We  showed  that   in  two  equivalent
expressions, one  deals with  the single nucleon  spectral function  as a
remnant of the nuclear response and the response of a moving nucleon.  In
those expressions, the  nucleon is either on, or off  its mass shell.  No
matter  of  principle is  involved,  since  the  two expressions  can  be
transformed one into the other.

Our  results are  sufficiently  general to  encompass  all previous
formulations. In particular Eqs. (\ref{a41}), (\ref{a42}) and
(\ref{a44}) are the precise formulations of the PWIA, without neglect
of off-shell  behavior in the  response for  the moving
nucleon, where  required.
We refer  in particular  to a  recent preprint  by Melnitchouk
$et\, al$ \cite{mel}  who study in detail deep  inelastic scattering from
an off-mass shell nucleon under general circumstances.  The above authors
assume that the nucleon  response is not {\mbox{\boldmath $P$}}-dependent
and perform the {\mbox{\boldmath $P$}}-integral on the remaining spectral
function.  As shown above, no such  assumption is necessary  and
Eqs. (\ref{a44}) provide  an  alternative, where  the P-integration  has
formally be performed.

Finally, as an alternative to the PWIA  for the nuclear part of the total
structure function,  we considered  a modification where  the knocked-out
particle is  not on its  mass-shell as in the  PWIA, and has  instead the
same off-shell mass as before  photon absorption.  In this approximation,
one retains at  least some FSI between knocked-out nucleon  and the core.
The  response in  the  PWIA  and the  above  modification  have the  same
asymptotic limit (\ref{a41}), but are quite different  for finite  $Q^2$
\cite{gurv,asr}.

Till today only sporadic attempts have been made to go beyond the PWIA.
Maybe the simplest reflection of FSI
is the replacement $M\to M_{eff}$, and which has for instance been
worked-out in \cite{mel}. A largely phenomenological treatment for finite
$A$ has been given by Frankfort et. al. \cite{fr}. Calculations
for nuclear matter related to relatively large $Q^2 \,\,A(e,e')X$
reactions (and not the EMC effect) and which
are based on nuclear dynamics are described in
\cite{asr,omar}. With emphasis in the past on the nuclear binding aspect
of calculations of the EMC effect, it is clearly of interest to compare
those with feasible determinations of FSI. Those calcuations are
currently in progress  and will be reported elsewhere.

\section{Acknowledgement.}

The authors would like to than Byron Jennings, Shim'on Levit and
Roland Rosenfelder for discussions on the subject matter.

{\bf Figure Captions.}

Fig. 1a,b,c.  Kinematics  in the description of the  asymptotic limits of
structure functions with tragets  at  rest,   corresponding   to
$W_{\alpha/{\rm  N}}$,  $W_{\rm  N/A}$  and  $W_{\alpha/{\rm  A}}$  (Eqs.
(\ref{a17a}),  (\ref{a17b})  and  (\ref{a18})).   Crosses  on  marks  the
particles, which are on shell in the GRS approach.

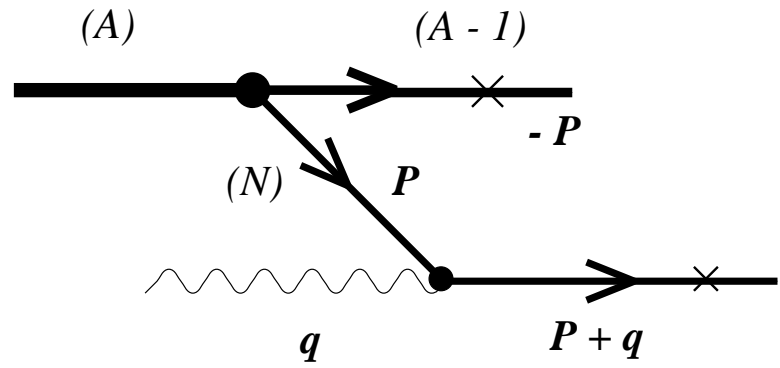
Fig. 2.  Kinematics for nuclear structure function $W_{\rm N/A}$ in PWIA.
Crosses mark on-shell particles.  Figure illustrates non-relativistic and
relativistic cases.

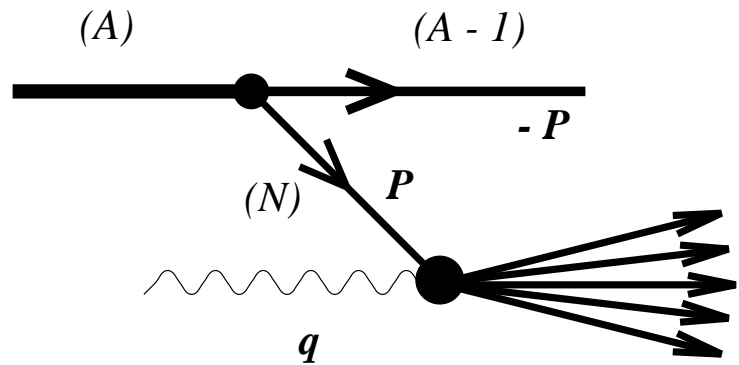
Fig.  3.    Amplitude,  underlying   total  nuclear   structure  function
$W_{\alpha/{\rm  A}}$  in  PWIA.   Photon vertex  induces  exact  nucleon
structure    function   $W_{\alpha/{\rm    N}}$.    Figure    illustrates
non-relativistic and relativistic cases.


\begin{references}
\bibitem{fnote} The sign convention in Eqs. (\ref{a30}) is in agreement
with  $y<0$, respectively $x>1$ for elastic scattering.
\end{references}
\end{document}